\documentclass[]{aa}  

\usepackage{graphicx}
\usepackage{amsmath}
\usepackage{txfonts}
\usepackage{gensymb}
\usepackage{natbib}
\bibpunct{(}{)}{;}{a}{}{,} % to follow the A&A style
\begin{document}

   \title{Experimental investigations of diacetylene ice photochemistry in Titan’s atmospheric conditions}

   \subtitle{}

   \author{Benjamin Fleury\inst{1,2} 
          \and Murthy S. Gudipati\inst{2}
          \and Isabelle Couturier-Tamburelli\inst{3}
          }

   \institute{Université Paris Cité and Univ Paris Est Creteil, CNRS, LISA, F-75013 Paris, France\\
              \email{benjamin.fleury@lisa.ipsl.fr}
        \and 
            Jet Propulsion Laboratory, California Institute of Technology, 4800 Oak Grove Drive, Pasadena, California 91109, USA
         \and
             Aix-Marseille Université, CNRS, PIIM, UMR 7345, 13013 Marseille, France\\
             \email{isabelle.couturier@univ-amu.fr}
             }

   \date{Received November 17, 2023; accepted January 31, 2024}

% \abstract{}{}{}{}{} 
% 5 {} token are mandatory
 
  \abstract
   {A large fraction of the organic species produced photochemically in the atmosphere of Titan can condense to form ice particles in the stratosphere and in the troposphere. According to various studies, diacetylene (C$_{4}$H$_{2}$) condenses below 100~km where it can be exposed to ultraviolet radiation.}
   {We studied experimentally the photochemistry of diacetylene ice (C$_{4}$H$_{2}$) to evaluate its potential role in the lower altitude photochemistry of Titan’s atmospheric ices.}
   {C$_{4}$H$_{2}$ ice films were irradiated with near-ultraviolet (near-UV) photons ($\lambda$ > 300~nm) with different UV sources to assess the impact of the wavelengths of photons on the photochemistry of C$_{4}$H$_{2}$. The evolution of the ice's composition was monitored using spectroscopic techniques.}
   {Our results reveal that diacetylene ice is reactive through singlet-triplet absorption, similar to the photochemistry of other organic ices of Titan (such as dicyanoacetylene C$_{4}$N$_{2}$ ice) that we investigated previously. Several chemical processes occurred during the photolysis: the hydrogenation of C$_{4}$H$_{2}$ to form other C$_{4}$ hydrocarbons (vinylacetylene C$_{4}$H$_{4}$ to butane C$_{4}$H$_{10}$); the formation of larger and highly polymerizable hydrocarbons, such as triacetylene (C$_{6}$H$_{2}$); and the formation of an organic polymer that is  stable at room temperature.}
   {The nondetection of diacetylene ice in Titan's atmosphere or surface could be rationalized based on our experimental results that C$_{4}$H$_{2}$ is photochemically highly reactive in the solid phase when exposed to near-UV radiation  that reaches Titan's lower altitudes and surface. C$_{4}$H$_{2}$ may be one of the key molecules promoting the chemistry in the ices and aerosols of Titan’s haze layers, especially in the case of co-condensation with other organic volatiles, with which it could initiate more complex solid-phase chemistry.}

   \keywords{Astrochemistry -- Methods: laboratory: molecular -- Techniques: spectroscopic -- Planets and satellites: atmospheres -- Ultraviolet: planetary systems
               }
    \titlerunning{Experimental Investigations of Diacetylene Ice Photochemistry}
    \maketitle
%
%-------------------------------------------------------------------

\section{Introduction}\label{Introduction}

The complex organic chemistry initiated in Titan’s thermosphere and ionosphere by solar photons and energetic particles of Saturn’s magnetosphere leads to the production of many volatile organics as well as organic aerosols \citep{Lavvasetal2013, Liangetal2007, Waiteetal.2007}. The decrease in temperature from the stratosphere to the tropopause \citep{Fulchignonietal.2005} allows the direct condensation of many volatiles (hydrocarbons and nitriles), thus forming ice particles or ice accreted on organic aerosols \citep{Andersonetal.2016, Barthetal.2003, Barthetal.2017, deKoketal.2008, Frereetal.1990, Lavvasetal2011, Mayoetal2005, Raulinetal2002, Saganetal1984, Samuelsonetal.1991, Samuelsonetal.1997}.

Gas molecules and organic hazes in Titan’s atmosphere significantly attenuate high-energy photons. As a result, only longer-wavelength photons make it through the stratosphere and the troposphere and eventually to the surface of Titan. \citet{Vuitonetal2019} model the flux of photons from 1.5 to 300~nm as a function of the altitude in the atmosphere of Titan. According to their work, photons with wavelengths shorter than 150~nm do not penetrate below $\sim$200~km, while photons in the 150 to 200~nm range are progressively absorbed from $\sim$200~km to $\sim$100~km. Finally, only photons in the 200 to 300~nm range penetrate below 100~km, down to $\sim$50~km. Our recent experimental studies have demonstrated that these longer-wavelength (200 nm and above) ultraviolet (UV) photons could efficiently drive the photochemistry of ices, resulting in the production of organic polymers \citep{Couturier-Tamburellietal.2014, Couturier-Tamburellietal.2015, Gudipatietal.2013, Mouzayetal2021a, Mouzayetal2021b} and to the photochemically driven incorporation of these molecules into existing haze particles \citep{Couturier-Tamburellietal.2018, Fleuryetal.2019}.

Our previous studies of pure-ice photochemistry in Titan’s atmosphere were focused on acetylene (C$_{2}$H$_{2}$), benzene (C$_{6}$H$_{6}$), and some nitriles (hydrogen cyanide HCN, cyanoacetylene HC$_{3}$N, dicyanoacetylene C$_{4}$N$_{2}$, and cyanodiacetylene HC$_{5}$N), as these species represent the composition of most identified ice clouds \citep{Andersonetal.2011, Andersonetal.2010, Andersonetal.2018b, deKoketal.2014, Khannaetal1987, Samuelsonetal.1997}. However, higher hydrocarbons may also condense in Titan’s stratosphere, as highlighted by the recent identification of ice clouds that contain benzene at Titan’s south pole \citep{Vinatieretal.2018} and by the discovery by the Cassini Composite InfraRed Spectrometer (CIRS) of the High Altitude South Polar (HASP) cloud \citep{Andersonetal.2018b} whose spectral signature is consistent with co-condensed HCN-C$_{6}$H$_{6}$ ices \citep{Andersonetal.2018b}. After C$_{6}$H$_{6}$, which  we studied previously \citep{Mouzayetal2021a, Mouzayetal2021b}, diacetylene (C$_{4}$H$_{2}$) is the next hydrocarbon to condense in the stratosphere of Titan at altitudes ranging from 70 to 130~km as a function of the latitude and the season according to different studies \citep{Andersonetal.2018b, Andersonetal.2016, Barthetal.2017}. Intrigued by the lack of detection of condensed C$_{4}$H$_{2}$ either in the atmosphere or at the surface of Titan, we decided to focus our study on the photochemistry of C$_{4}$H$_{2}$ ice as a potential sink for condensed C$_{4}$H$_{2}$ in these environments.

In Titan’s atmosphere, C$_{4}$H$_{2}$ is formed in the gas phase below 1200~km by the reaction of acetylene with C$_{2}$H radicals \citep{Chastaingetal.1998, Kaiseretal.2002}, while another mechanism involving the reaction of C$_{4}$H$_{3}$ with CH$_{3}$ radicals exists at higher altitude \citep{Vuitonetal2019}. Its mixing ratio varies significantly with altitude, decreasing from $\sim$ 7~$\times$~10$^{-6}$ at 981~km in the ionosphere \citep{Cuietal.2009} to~1~$\times$~10$^{-9}$ in the stratosphere \citep{Coustenisetal.2007, Coustenisetal.2010, Vinatieretal.2007, Vinatieretal.2010}. In addition, the atmospheric mixing ratio of diacetylene is subject to seasonal variations, leading to a strong polar enrichment in the winter hemisphere \citep{Coustenisetal.2016, Sylvestreetal.2018}. Further, among the major unsaturated hydrocarbons, C$_{4}$H$_{2}$ has one of the most extended UV absorptions toward long wavelengths involving the S$_{0}$$\rightarrow$S$_{1}$ (286~nm) and S$_{0}$$\rightarrow$T$_{1}$ (387~nm) excitations \citep{Fischeretal.2003, Vilaetal.2000} in addition to its strong photoabsorption cross section between 120 and 220~nm \citep{Ferradazetal.2009}. Our previous works have shown that direct photoexcitation from the singlet ground state to a triplet excited state is possible in the condensed phase \citep{Couturier-Tamburellietal.2014, Gudipatietal.2013}. Therefore, diacetylene is an excellent candidate for solid-state photochemistry and a possible source for the formation of larger volatile molecules or organic polymers in the atmosphere of Titan. In addition, such a reactive photochemical sink could also potentially explain the lack of C$_{4}$H$_{2}$-ice detection, whereas other ices such as HC$_{3}$N have been detected in Titan’s atmosphere \citep{Andersonetal.2010, Andersonetal.2016}. C$_{4}$H$_{2}$ is known to be photochemically reactive in the gas phase \citep{Bandyetal.1992a, Bandyetal.1992b, Delpechetal.1994, Frostetal.1995, Glickeretal.1987}, producing long carbon-chain species such as tri-acetylene (C$_{6}$H$_{2}$) and tetra-acetylene (C$_{8}$H$_{2}$). Therefore, C$_{4}$H$_{2}$ is suspected to be an important intermediate in the reaction pathways for the formation of large organic molecules that make up Titan’s atmospheric haze \citep{Abplanalpetal.2019, Kaiseretal.2012, Wilsonetal.2004}. Another possibility is that the nondetection of C$_{4}$H$_{2}$ ice in Titan’s atmosphere could be due to the existence of C$_{4}$H$_{2}$ co-condensate with another species such as HCN, that results in a modification of its spectral signature, as has been observed for the HC$_{3}$N-HCN or C$_{4}$N$_{2}$-HCN co-condensate \citep{Andersonetal.2018a}.

The work reported here presents the first experimental study of the photochemistry of diacetylene ice irradiated with near-UV ($\lambda >$ 300~nm) photons. We studied the role of longer-wavelength UV-irradiation in the chemical evolution of C$_{4}$H$_{2}$ ices, the formation of volatile species, and organic polymers in the lower atmosphere of Titan. Finally, we assess the influence of the UV source (monochromatic vs broadband) on the simulated photochemistry.

%--------------------------------------------------------------------
\section{Experimental methods and analysis protocols}\label{Experimental methods and analysis protocols}

%--------------------------------------------------------------------
\subsection{The TOAST and RING experimental setups}\label{The TOAST and RING experimental setups}

\begin{figure}[h]
    \centering
    \includegraphics [trim=0 150 0 120,clip,width=\hsize] {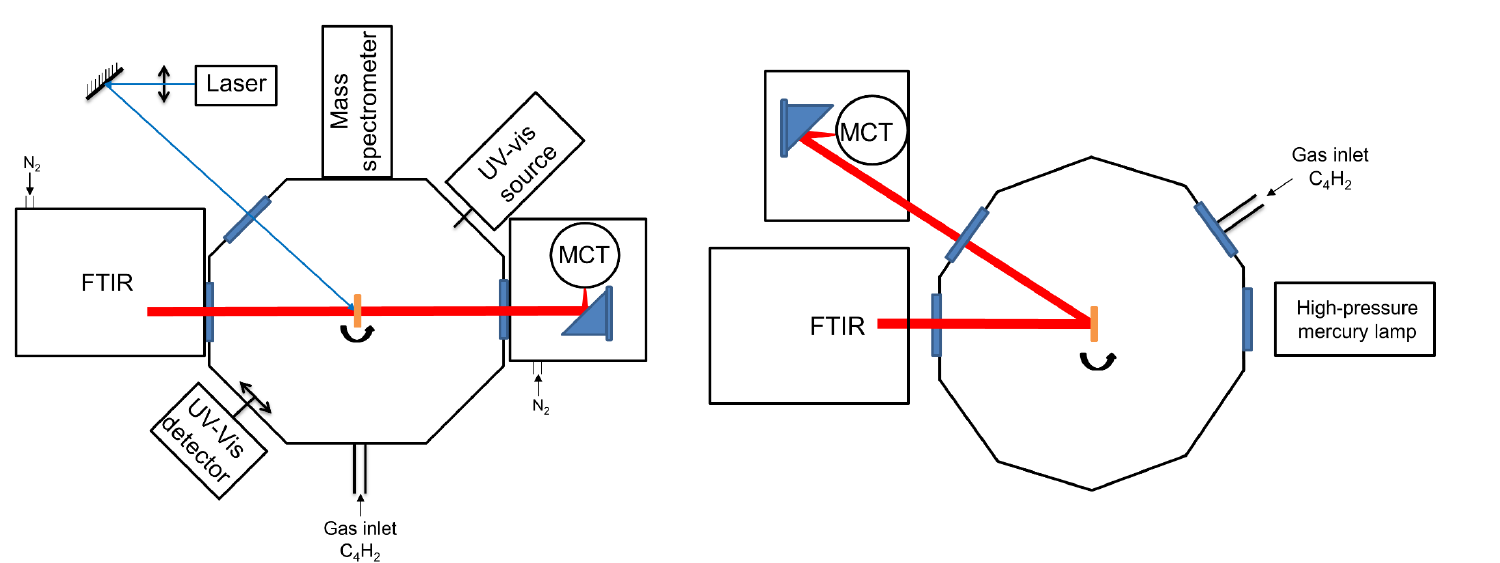}
    \caption {Schemes of the TOAST experimental setup at JPL (left) and the RING experimental setup at PIIM (right).}
    \label{Figure 1}%
\end{figure}

For this study we used the Titan Organic Aerosol SpecTroscopy (TOAST) experimental setup at the Jet Propulsion Laboratory, USA \citep{Fleuryetal.2019, Gudipatietal.2013}, and the RING experimental setup at the PIIM Laboratory at Aix-Marseille University, France \citep{Theuleetal.2011}. Figure \ref{Figure 1} presents the schemes of the TOAST and RING setups. We decided to use these two experimental setups to obtain complementary datasets. As described in greater detail below, the two setups used infrared (IR) spectroscopy in transmission and reflection modes as well as complementary wavelength ranges. In addition, the experimental setup used at JPL used a monochromatic wavelength for the ice irradiation, while the setup at PIIM used a polychromatic UV source, which allowed us to study the possible differences in the chemistry between monochromatic and polychromatic irradiation.

Diacetylene is not available commercially and needs to be synthesized. For this study diacetylene was synthesized according to the method described in the literature \citep{Khlifietal1995}, by the dropwise addition of 1,4 dichlroro-2-butyne (C$_{4}$H$_{4}$Cl$_{2}$) to a mixture of potassium hydroxide (KOH), dimethyl sulfoxide (DMSO), and water maintained at a temperature of 72~$\degree$C under nitrogen flow atmosphere. DMSO serves to increase the solubility of C$_{4}$H$_{2}$ in the aqueous phase. At the end of the addition the temperature was increased to 95~$\degree$C and held at this temperature for 15~min. C$_{4}$H$_{2}$ gas was then driven off by bubbling N$_{2}$ gas through the liquid, passed through a calcium chloride (CaCl$_{2}$) trap to remove traces of water, and condensed in a trap cooled with liquid nitrogen (LN$_{2}$). Finally, C$_{4}$H$_{2}$ was purified by vacuum distillation and its purity was checked using mass spectrometry. 

At JPL, ice films were formed by vapor deposition on a sapphire window fixed on the sample holder of a closed-cycle helium cryostat, located within a high vacuum chamber with a background pressure $\sim$8~$\times$~10$^{-9}$ mbar. The cryostat has a resistive heater and a temperature controller to precisely control the temperature of the sample from 10 to 300~K. The temperature of the sample holder was measured using a silicon diode. In this study, C$_{4}$H$_{2}$ vapor was deposited from a gas line at a pressure of 1~$\times$~10$^{-6}$~mbar for 15~min on a sapphire window cooled  to 70~K. This is lower than the temperature at which C$_{4}$H$_{2}$ could condense in Titan's atmosphere which is $\sim$110~K for an altitude of $\sim$70~km \citep{Barthetal.2017, Lavvasetal2011}. However, we made this choice based on different experimental constraints. First, a previous experimental study determined that diacetylene starts to sublimate at $\sim$100~K \citep{Zhouetal.2009} under a background pressure of $\sim$5~$\times$~10$^{-9}$~mbar, similar to the pressures reached in our experimental chamber (i.e., $\sim$8~$\times$~10$^{-9}$~mbar). It was therefore necessary to use a temperature lower than 100~K. We decided to use 70~K, a temperature  much lower than 100~K, to avoid a possible slow sublimation of C$_{4}$H$_{2}$ during the irradiation experiments, which typically took several hours to complete. Sublimation could happen at any temperature in a high vacuum chamber because an ice is in equilibrium with its vapor phase. As the experimental chambers have turbopumps, vapor can be pumped, inducing more and more sublimation of the ice over time. Nevertheless, the lower the temperature away from the sublimation temperature ($\sim$100~K), the weaker the ice's sublimation, and it was negligible at 70~K for the duration of the trial (which we experimentally determined). 

The prepared ice film, about a micrometer thick (see Sect. \ref{Spectroscopy of C4H2 ice deposited at 70 K}), was subsequently irradiated for 6~hr at 1~hr doses, with photons at 355~nm obtained by a third harmonic generation from a nanosecond pulsed Nd:YAG laser (Quantel) with a repetition rate of 20~Hz. After each hour of irradiation, the spectra were measured before continuing the photolysis. The diameter of the initial circular laser beam was expanded from 3 to 20~mm using a plano-convex lens (75~mm focal length) that reduced the laser photons flux by a factor of 44 to uniformly illuminate the entire surface of the ice film and to avoid multi-photon processes. In a previous publication we showed that photons at 355~nm produced by a defocused nanosecond pulsed laser drove solid-phase photochemistry via a single-photon process \citep{Gudipatietal.2013}, and one-photon absorption was shown to dominate even at 8~mJ/pulse or 9.1~$\times$~10$^{16}$~photons~cm$^{-2}$~s$^{-1}$ at 355~nm and 20~Hz repetition rate. The present work was carried out under similar conditions: at the same wavelength at 20~Hz repetition rate with an energy of 10.5~mJ/pulse or 1.3~$\times$~10$^{17}$~photons~cm$^{-2}$~s$^{-1}$. Pulse energies were measured at the laser outlet, while photons fluxes were calculated at the sample's surface after expansion of the laser. The evolution of the diacetylene ice film was monitored after each irradiation using transmission-absorption IR and ultraviolet-visible (UV-Vis) spectroscopy. After irradiation, the sample was warmed to 180~K under vacuum using temperature-programmed desorption (TPD) at 1~K~min$^{-1}$ in order to sublimate the photo-produced volatiles. During the experiment, the composition of the sublimated gas-phase molecules was also monitored using in situ mass spectrometry. 

At PIIM, ice films were formed by vapor deposition on a copper substrate fixed on the sample holder of a closed-cycle helium cryostat, located within a high vacuum chamber with a background pressure of $\sim$8~$\times$~10$^{-9}$~mbar. The cryostat has a resistive heater and a temperature controller to precisely control the temperature of the sample from 10~K to 300~K. The temperature of the sample holder was measured using a silicon diode. We deposited 0.6~mbar of gas phase C$_{4}$H$_{2}$ for around 20~min on the copper surface kept at 70~K at a pressure of 1~$\times$~10$^{-6}$~mbar. In contrast to the JPL experiments where the ice was irradiated with a single wavelength laser, during the RING experiments, a diacetylene ice film (a few hundred nanometers thick; see Sect. \ref{Spectroscopy of C4H2 ice deposited at 70 K}) was irradiated with broadband UV photons ($\lambda >$ 300~nm) generated by an Osram 200~W high-pressure mercury lamp equipped with a cutoff filter at 300~nm. The total photon flux of the lamp at wavelengths above 300~nm has been estimated to be about 2.75~$\times$~10$^{16}$~photons~cm$^{-2}$~s$^{-1}$ \citep{Couturier-Tamburellietal.2014}. After about 5~hr of irradiation, the ice samples were slowly warmed up to room temperature to evaporate the volatile species. Changes in the ice samples were monitored by reflection IR spectroscopy.

%-----------------------------------------------------------------
\subsection{IR spectroscopy}\label{Infrared spectroscopy}

At JPL, IR spectra were recorded in transmission with a Thermo Scientific Nicolet 6700 Fourier Transform InfraRed (FTIR) spectrometer. After passing through the sample, the IR beam was focused on a mercury cadmium telluride (MCT) detector cooled to 77~K with LN$_{2}$. Each spectrum consists of an average of 200 scans with a resolution of 1~cm$^{-1}$ and covers the 1600~cm$^{-1}$ (sapphire cut on) to 4500~cm$^{-1}$ range.
Reflection absorption IR spectra were measured at PIIM from 4000 to 600~cm$^{-1}$ using a Vertex 70 FTIR spectrometer. Each spectrum was averaged over 100 scans with a resolution of 1~cm$^{-1}$. At the copper surface the IR beam was reflected with a 36$\degree$ angle and focused on a MCT detector cooled to 77~K. 
The two experiments at PIIM and JPL are therefore complementary in terms of spectral ranges, and provide a large spectral covering from 4500 to 600~cm$^{-1}$. The IR spectra recorded at JPL and PIIM were used to calculate the thickness of the deposited ice films and the diacetylene column density (see Sect. \ref{Results}). To do this, we used the Beer-Lambert law. In the case of the reflection absorption spectra recorded at PIIM, this is an approximation that could bias the calculated values. Even so, this approach has been  used in experimental studies using reflection absorption spectroscopy \citep{Bennett_2005,Couturier-Tamburellietal.2018, Mouzayetal2021b, Nobleetal2012}. To take this problem into account and limit the error on the calculated values in the case of the PIIM experiments, we decided to divide the integrated absorbance in the equations used in Sect. \ref{Results} by a factor of 2 to correct for the incoming and outgoing infrared beam through the ice film.

%-----------------------------------------------------------------
\subsection{UV-Vis spectroscopy}\label{UV-Vis spectroscopy}

At JPL, the UV-Vis spectra were measured from 210 to 1100~nm with an incidence angle of 45$\degree$ (Fig. \ref{Figure 1}), using an Ocean Optics DH-2000 deuterium-halogen lamp and a USB4000 spectrograph connected to the chamber via optical fiber feedthrough. Each spectrum was averaged over 500 scans with 20~ms of integration per scan.

%-----------------------------------------------------------------
\subsection{Mass spectrometry}\label{Mass spectrometry}

At JPL, the analysis by mass spectrometry of the molecules desorbed during the TPD was achieved with a quadrupole mass spectrometer (RGA 300, SRS instrument), with a mass  range of  1 to 300~a.m.u. and equipped with a continuous dynode electron multiplier (CDEM) detector, enabling measurement of trace molecules at partial pressures as low as 1~$\times$~10$^{-12}$~mbar. The molecules were ionized by electron impact with an energy of 70~eV. The mass resolution (m/$\Delta$m) is 100 at \textit{m/z} 100. 

%-----------------------------------------------------------------
\section{Results}\label{Results}

%-----------------------------------------------------------------
\subsection{Spectroscopy of C$_{4}$H$_{2}$ ice deposited at 70~K}\label{Spectroscopy of C4H2 ice deposited at 70 K}

\begin{figure} [h]
    \centering
        \includegraphics[width=0.48\textwidth, clip]{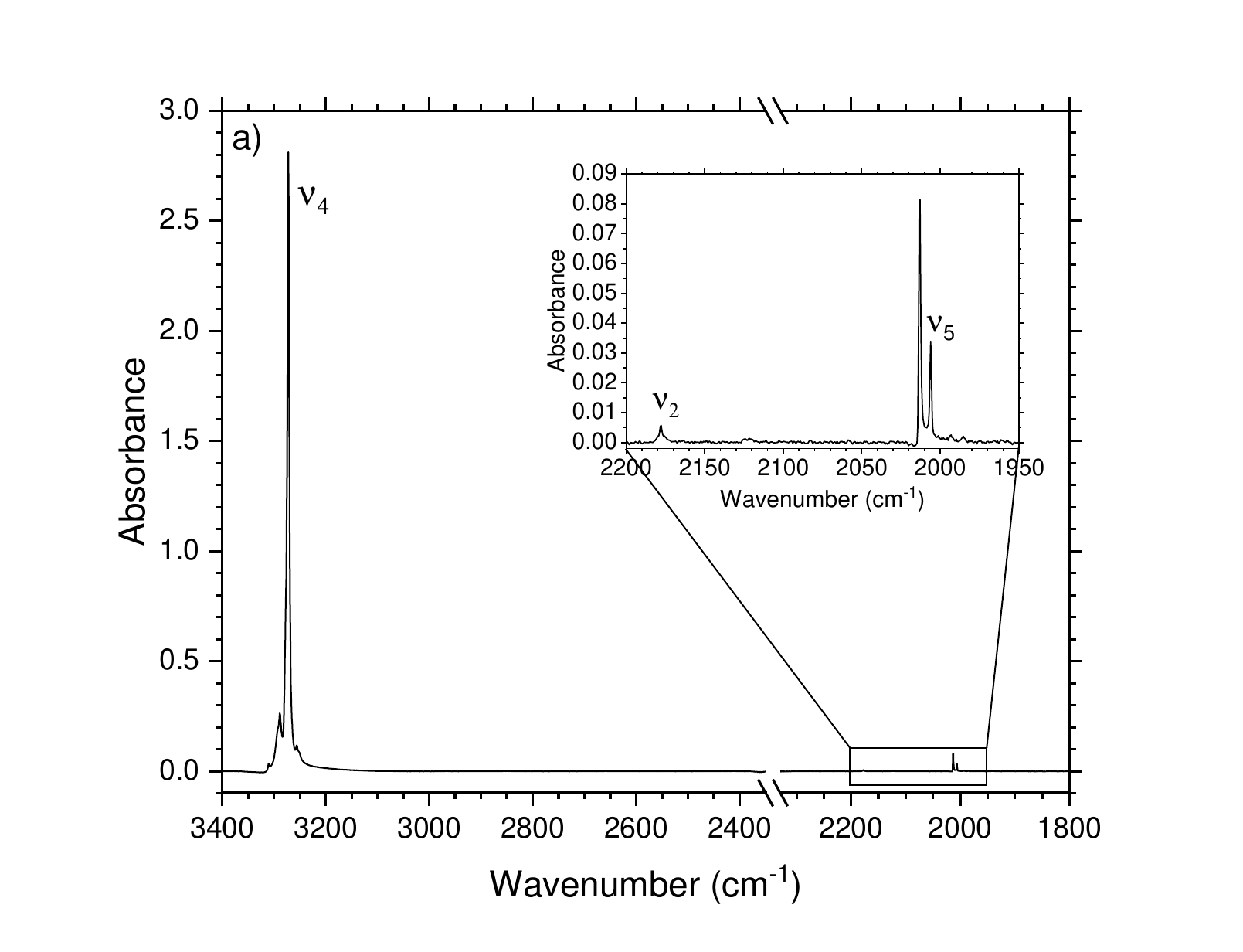}
        \includegraphics[width=0.48\textwidth, clip]{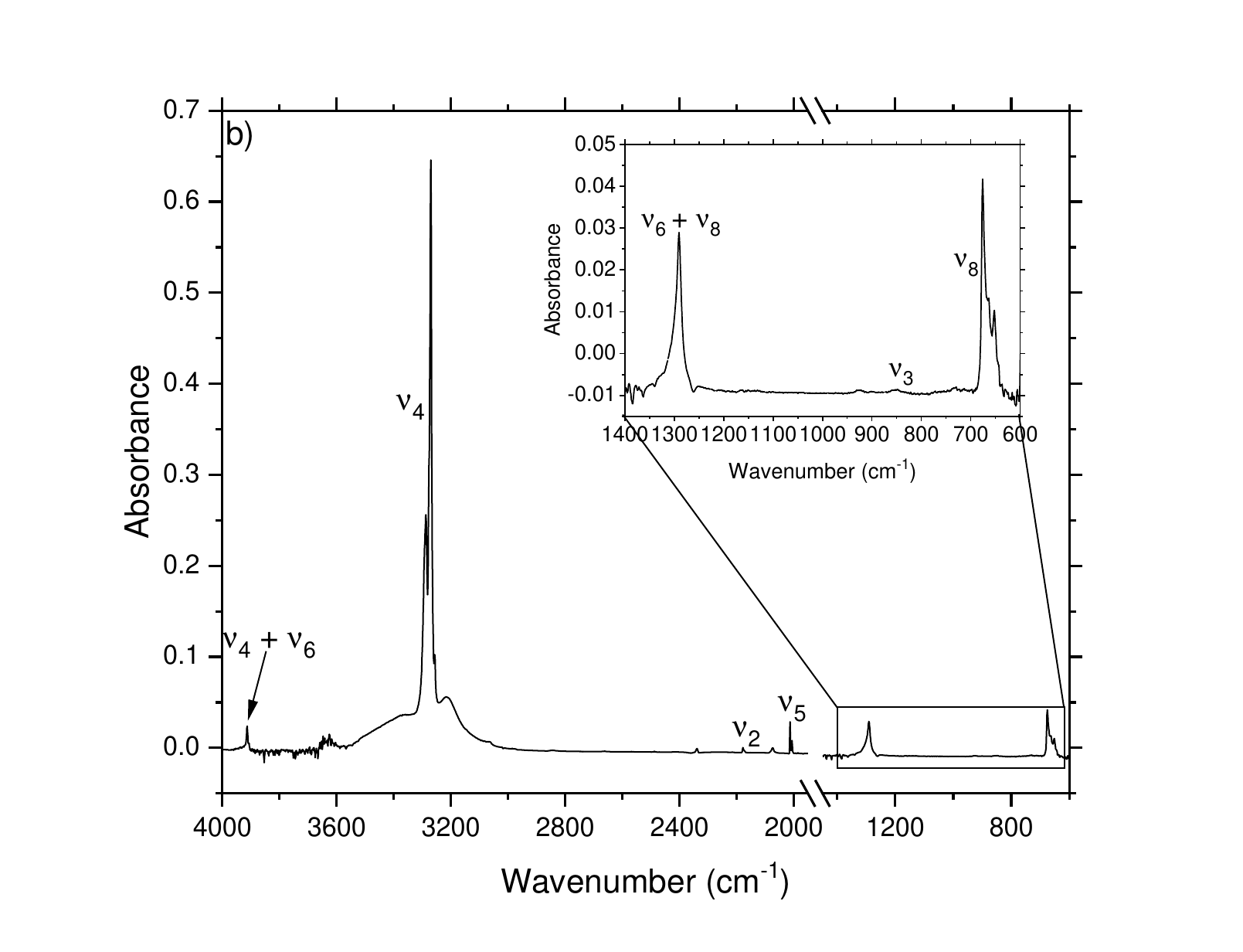}
    \caption {Infrared spectra of C$_{4}$H$_{2}$ films obtained at JPL and PIIM independently. (a) IR spectrum of C$_{4}$H$_{2}$ ice deposited at 70~K at JPL  on a sapphire window. The splitting of the $\nu_{5}$ band is due to the crystalline nature of the ice film. The spectrum is cropped at 2340~cm$^{-1}$ to remove the absorption band due to gas-phase CO$_{2}$ in the path of the IR beam outside the vacuum chamber. (b) IR spectrum of C$_{4}$H$_{2}$ ice deposited at 70~K at PIIM on a copper substrate.}\label{Figure 2}
\end{figure}

Figure \ref{Figure 2} presents the IR spectra of the diacetylene ice films deposited at 70~K at JPL and PIIM in the 3400-1800~cm$^{-1}$ region and in the 4000-600~cm$^{-1}$ region, respectively. Three fundamental absorption bands of C$_{4}$H$_{2}$ are observed in the JPL spectrum, in agreement with previous studies \citep{Khannaetal.1988, Zhouetal.2009}. The most intense peak at 3272~cm$^{-1}$ is assigned to the $\nu_{4}$ asymmetric stretching mode of the C-H bond. This band is saturated on the JPL spectrum because of the thick ice film used for the experiment (see discussion below). Weaker absorption bands at 2013 and 2177~cm$^{-1}$ are assigned to the $\nu_{5}$ asymmetric stretching mode and the $\nu_{2}$ symmetric stretching mode of the C$\equiv$C bond, respectively. Two other fundamental absorption bands of C$_{4}$H$_{2}$ are observed in the PIIM's spectrum located at 674 and 847~cm$^{-1}$ and assigned to the $\nu_{8}$ CH bending mode and to the $\nu_{3}$ stretching mode of the C-C bond respectively. The position and shape of the absorption bands suggest that C$_{4}$H$_{2}$ is in a crystalline form when deposited at 70~K. This agrees with the previous experimental study of \citet{Zhouetal.2009}, which found that for C$_{4}$H$_{2}$ ice the amorphous to crystalline transition occurs around 70~K.

The diacetylene ice thickness h can be calculated as 
\begin{equation}
    h = \frac{\int \tau_{\rm \nu} \rm d\nu}{\int \alpha \rm d\nu}
    \label{Eq.1}
,\end{equation}
where \emph{$\int$$\tau_{\nu}$\rm d$\nu$} is the integrated optical depth of the band, which was determined from the absorbance of the band, and \emph{$\int$$\alpha$\rm d$\nu$} is the integrated extinction coefficient of the band (cm$^{-1}$). For our calculation, we used an integrated extinction coefficient value of 3.9~$\times$~10$^{3}$~cm$^{-1}$ for the 2013~cm$^{-1}$ band, determined for crystalline diacetylene at 70~K by \citet{Khannaetal.1988}. We calculated a diacetylene ice thickness of $\sim$374~nm for the PIIM experiment and$\sim$1.1 µm for the JPL experiment, reflecting the difference in the deposition conditions between experiments carried out at JPL and PIIM. Although the thicker deposit used at JPL leads to a saturation of the C-H absorption band of C$_{4}$H$_{2}$ (see Fig. \ref{Figure 2}), we decided to proceed with this ice thickness to enhance the formation of volatile species and increase their detectability by IR and UV spectroscopy and mass spectrometry.

\begin{figure}[h]
    \centering
    \includegraphics [width=\hsize] {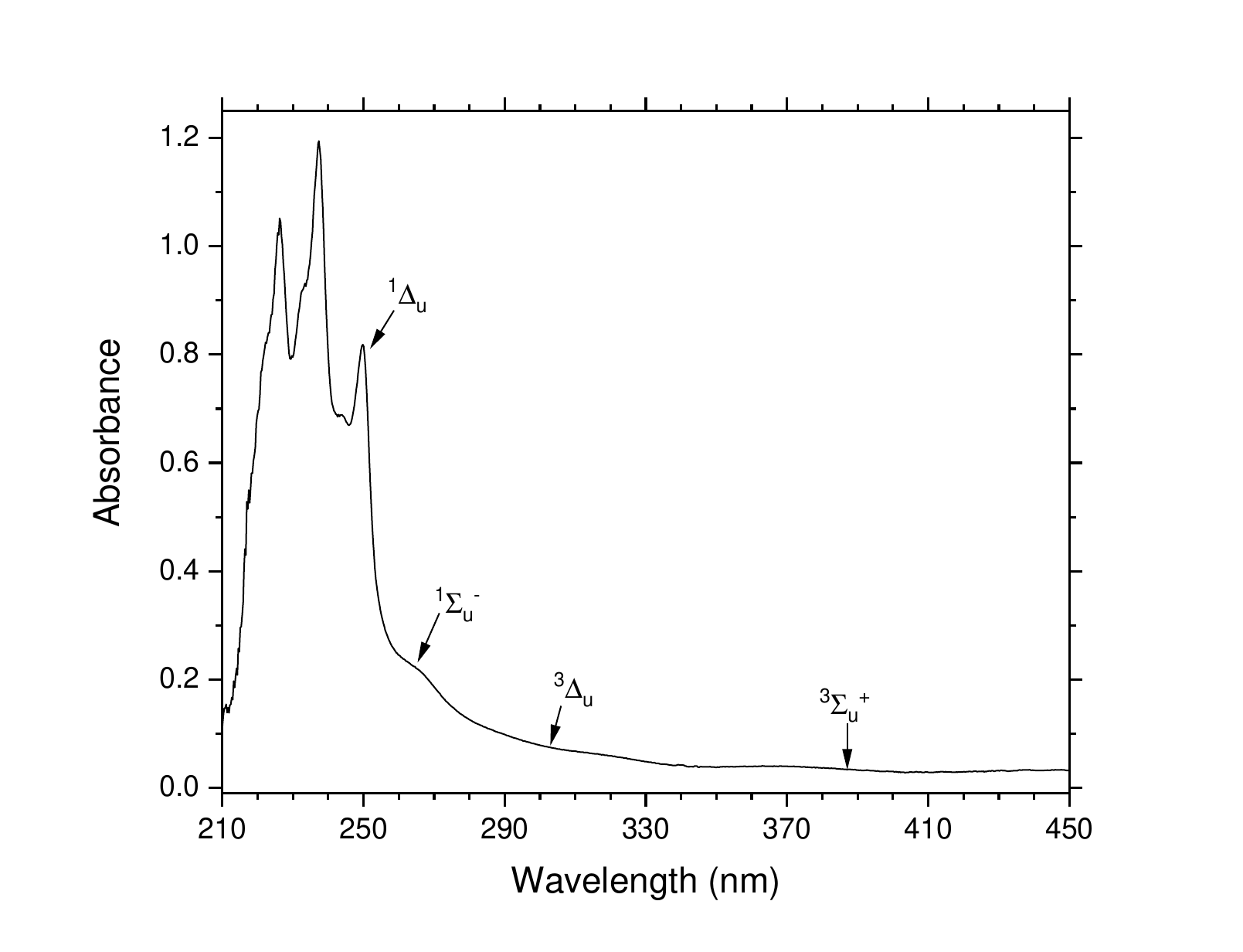}
    \caption {UV spectrum of C$_{4}$H$_{2}$ ice deposited at 70~K at JPL on a sapphire window. The positions of the center of some electronic transitions of C$_{4}$H$_{2}$ are indicated (see discussion in Sect. \ref{Chemical pathways for hydrocarbons formations}).}
    \label{Figure 3}%
\end{figure}

As presented in Fig. \ref{Figure 3}, C$_{4}$H$_{2}$ ice at 70~K has strong absorption in the UV, with structured absorption bands at 226, 237, and 249~nm. These bands are separated by 2040~cm$^{-1}$ wavenumbers, corresponding to $\nu_{5}$ vibronic coupling in the singlet exited state $^{1}\Delta_{u}$. These absorption bands are close to those measured in the gas phase for C$_{4}$H$_{2}$, but as expected they are shifted by $\sim$5~nm toward longer wavelengths compared to the gas-phase bands \citep{Jollyetal.2008, Smithetal1998}. However, in contrast to the gas phase, we observed in the solid phase the presence of a weaker absorption continuum from 250 to 400~nm.

%-----------------------------------------------------------------
\subsection{Photochemistry of pure C$_{4}$H$_{2}$ ice at $\lambda$ > 300 nm}\label{Photochemistry of pure C4H2 ice at λ > 300 nm}

\begin{figure*}[h]
    \centering
    \includegraphics [width=\hsize] {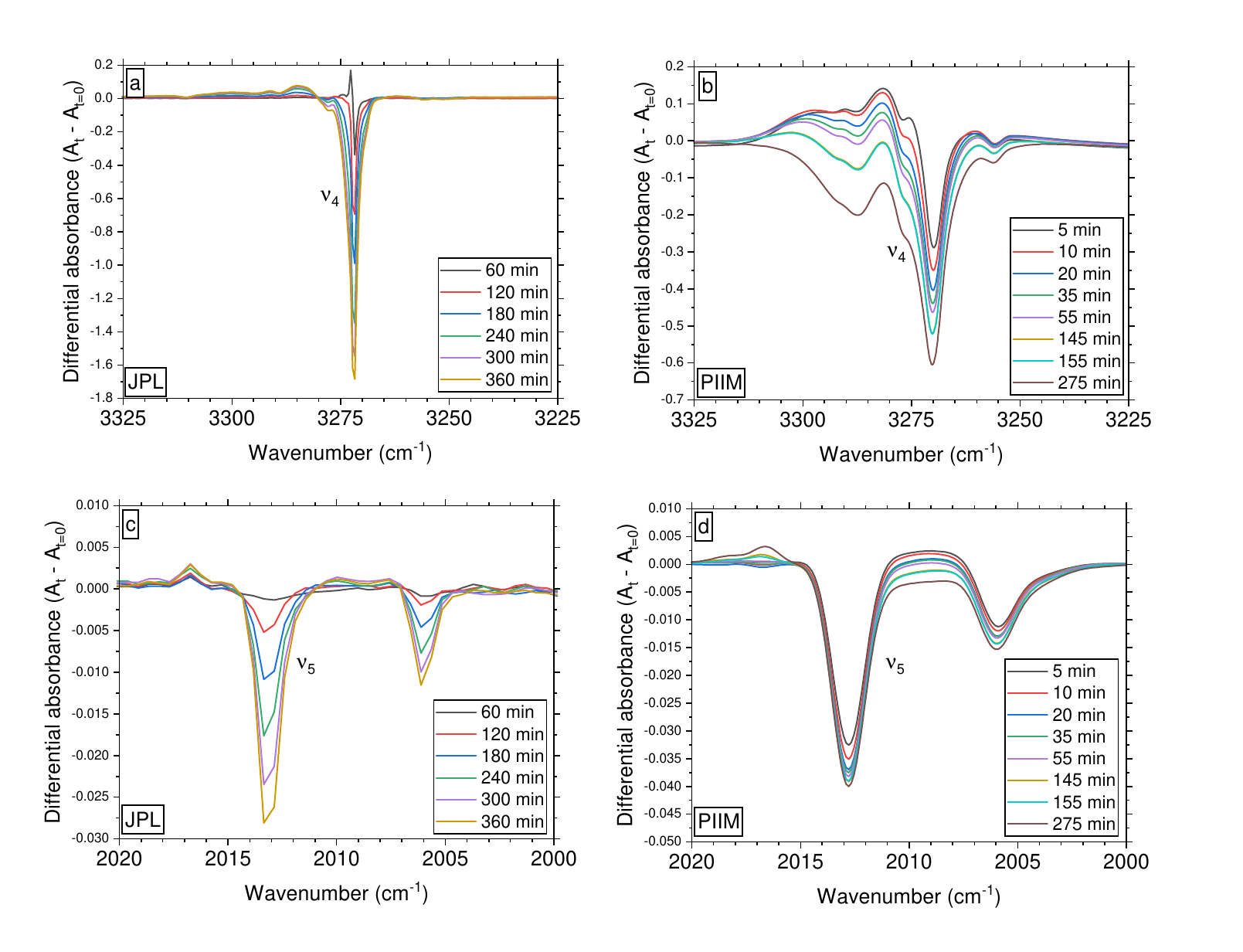}
    \caption {Changes to IR spectra observed during near-UV irradiation of C$_{4}$H$_{2}$ ice films. a) Evolution of the differential absorbance (A$_{t}$ – A$_{t=0}$) of the $\nu_{4}$ band of C$_{4}$H$_{2}$ as function of the time of irradiation at 355~nm at JPL. b) Evolution of the differential absorbance of the $\nu_{4}$ band of C$_{4}$H$_{2}$ as function of the time of irradiation at $\lambda$ > 300~nm at PIIM. c) Evolution of the differential absorbance of the $\nu_{5}$ band of C$_{4}$H$_{2}$ as function of the time of irradiation at 355~nm at JPL. d) Evolution of the differential absorbance of the $\nu_{5}$ band of C$_{4}$H$_{2}$ as function of the time of irradiation at $\lambda$ > 300~nm at PIIM.}
    \label{Figure 4}%
\end{figure*}

At JPL the C$_{4}$H$_{2}$ ice film was irradiated for 6~hr with a 355~nm defocused laser. We monitored the gas-phase volatiles in the chamber using mass spectrometry during the irradiation and did not observe any photodesorption of diacetylene. In addition, IR and UV-Vis spectra were recorded after each irradiation period. We  used the evolution of the diacetylene absorption bands at 3272 and 2013~cm$^{-1}$ after each irradiation period to monitor its consumption due to photochemical reactions. Figure \ref{Figure 4} (Panels A and C) presents the evolution of the differential absorbance of these two absorption bands of C$_{4}$H$_{2}$ after each irradiation period. The differential absorbance is the difference between the absorbance after a time “t” of irradiation A$_{t}$ and the initial absorbance A$_{t=0}$. After 1 hr of irradiation we observed a small change in the C-H absorption band at 3272~cm$^{-1}$. However, this change is not conclusive (increase and decrease of the band) because of the saturation of the C-H absorption band at the beginning of the experiment (absorbance value >> 2.5, Fig. \ref{Figure 2}). In addition, we observed a very small depletion of the $\nu_{5}$ band at 2013~cm$^{-1}$. However, when we pursued the irradiation further, the absorbance of both diacetylene bands decreased, indicating a significant depletion of the ice sample. We also observed an increase in the absorbance at higher wavenumber, with broad peaks at 3284 and 3291~cm$^{-1}$ as well as a larger continuum at higher wavenumber. This indicates the formation of new products from the C$_{4}$H$_{2}$ ice photochemistry. The peak at 3284~cm$^{-1}$ may be attributed to vinylacetylene, C$_{4}$H$_{4}$ \citep{Kimetal2009}. The same peak was observed in previous laboratory experiments involving acetylene ice photochemistry \citep{Cuylleetal.2014}.

The C$_{4}$H$_{2}$ ice film irradiation was performed at PIIM with a lamp generating photons at $\lambda$ > 300~nm in order to see if broadband irradiation leads to different results. Nevertheless, the results of these experiments are similar to those performed at JPL, as illustrated by the IR spectra shown in Fig. \ref{Figure 4} (Panels B and D). The subtraction spectra reveal the appearance of new infrared bands, listed in Table \ref{Table1}, in addition to those observed at 3284 and 3291~cm$^{-1}$ in the laser experiment. They are attributed without ambiguity to C$_{4}$H$_{4}$ when compared to the data of \citet{Kimetal2009}. From these results obtained at JPL and PIIM, we confirm that C$_{4}$H$_{4}$ is one of the photoproducts of C$_{4}$H$_{2}$ ice photochemistry. We note that the sapphire window used for transmission-absorption spectroscopy in the JPL experiments cuts off below 1600~cm$^{-1}$. For this reason, and based on the relative intensities of infrared bands obtained at PIIM, we observed only the $\nu_{1}$ band of C$_{4}$H$_{4}$ in the JPL experiments. On the other hand, the PIIM experimental setup  records the entire mid-infrared region thanks to the use of a reflection-absorption configuration. As a result, several absorption bands assigned to C$_{4}$H$_{4}$ are also observed in the experiments conducted at PIIM, increasing the confidence of our assignment. We also note that in the transmission-absorption mode employed at JPL, the sample is $\sim$1100~nm thick, which is also the IR beam pathlength. Under reflection-absorption conditions, the IR beam penetrates through longer than twice the ice thickness of 374~nm. For this reason, we could be sampling twice the amount of the ice. 

\begin{table*}[h]
    \centering
        \caption{Infrared absorption bands of C$_{4}$H$_{4}$ ice observed in our experimental conditions.}\label{Positions cm-1 and assignments of the IR absorption bands of C$_{4}$H$_{4}$ observed after our irradiation experiments. Relative intensities are given in italic.}
    \begin{tabular}{ccccc}
    \hline\hline
     Assignment & \citet{Kimetal2009} 100~K & JPL 70~K & PIIM 70~K & PIIM 70~K \\ %Heading lines
     & Position (cm$^{-1}$) & Position (cm$^{-1}$) & Position (cm$^{-1}$) & Relative intensity \\ %Heading lines
      \hline
      $\nu_{1}$ & 3288 & 3291/3284 & 3290/3281 & 100 \\
       $\nu_{3}$ & 3049 & & 3037 & 1.5 \\
       $\nu_{4}$ & 3014 & & 3017 & 2.1 \\
       $\nu_{6}$ + $\nu_{7}$ & & & 2965 & 0.5 \\
       $\nu_{5}$ & 2103 & & 2104 & 2.0 \\
       $\nu_{6}$ & 1599 & & 1592 & 0.9 \\
      $\nu_{10}$ + $\nu_{12}$ & 1374 & & 1384 & 10.3 \\
      2$\nu_{11}$/2$\nu_{17}$/ $\nu_{11}$ + $\nu_{17}$ & 1259 & & 1262 & 15.9  \\
      \hline
    \end{tabular}
    \tablefoot{The position of the absorption bands observed after irradiation in the JPL and PIIM experiments at 70~K is shown. The position of the IR absorption bands of crystalline C$_{4}$H$_{4}$ ice at 100~K determined experimentally by \citet{Kimetal2009} is also shown for reference. Relative intensities of the IR bands observed in the PIIM experiment were calculated from the integrated areas of the IR bands.}
    \label{Table1}
\end{table*}

\begin{figure}[h]
    \centering
    \includegraphics [width=\hsize] {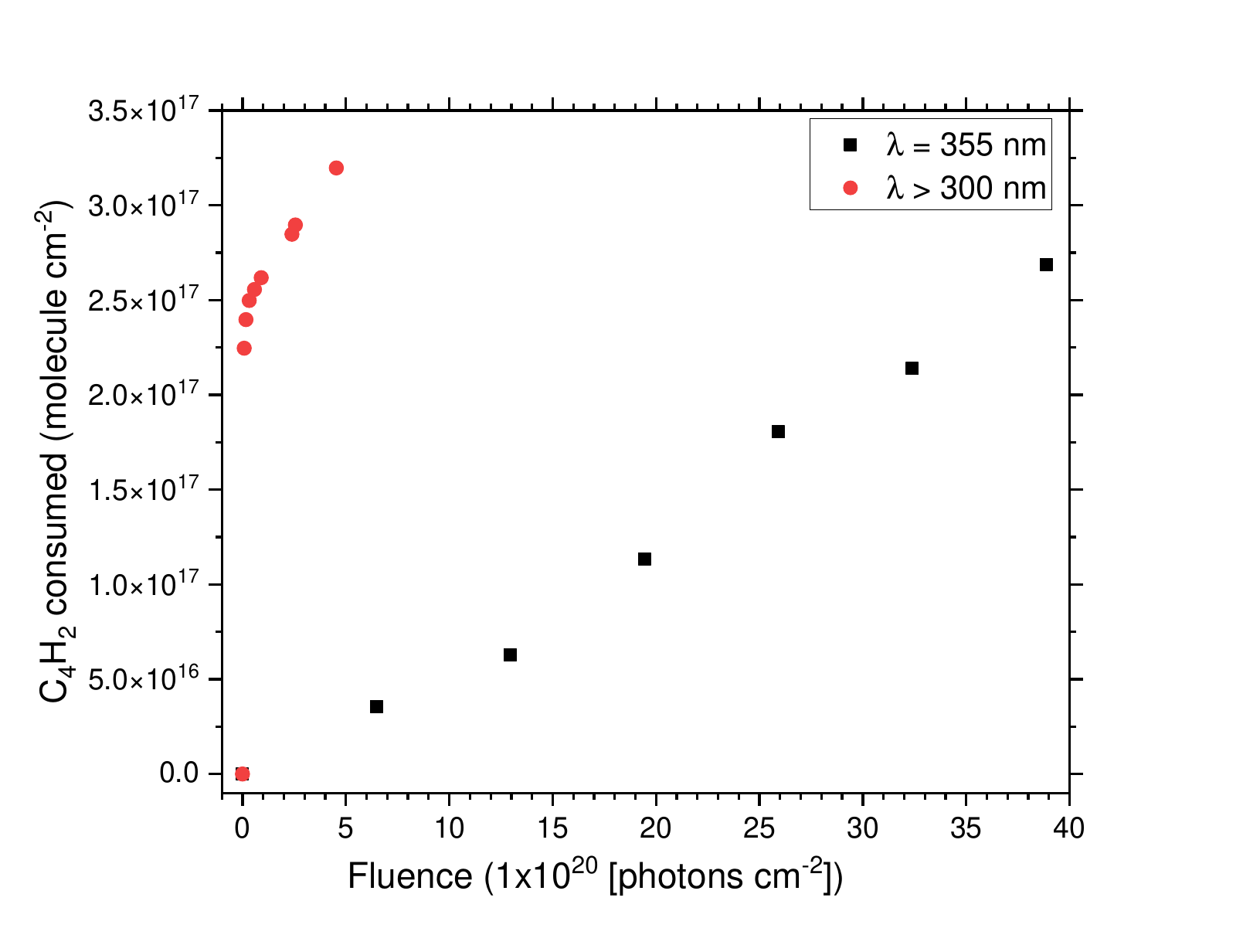}
    \caption {Quantity of C$_{4}$H$_{2}$ consumed as a function of the photon-fluence (JPL, 355~nm, black square; PIIM, $\lambda$ > 300~nm, red circle). The quantity is expressed in column density and was calculated from the IR spectra.}
    \label{Figure 5}%
\end{figure}

We derived the evolution of the C$_{4}$H$_{2}$ column density consumed as a function of the photon fluences for both experiments from the IR data shown in Fig. \ref{Figure 4}. We decided to calculate the column density using the integrated $\nu_{5}$ band area of C$_{4}$H$_{2}$ at 2013~cm$^{-1}$ because the other band at 3272~cm$^{-1}$ was saturated. The consumed column density can be calculated with Eq. \ref{Eq.2},
\begin{equation}
    N_{\rm c} = N_{\rm 0} - N_{\rm f}
    \label{Eq.2}
,\end{equation}
where \emph{N$_{\rm c}$} is the consumed column density (molecules~cm$^{-2}$) after a fluence f, \emph{N$_{\rm 0}$} is the initial column density (molecules~cm$^{-2}$) of C$_{4}$H$_{2}$, and \emph{N$_{\rm f}$} is the column density (molecules~cm$^{-2}$) of C$_{4}$H$_{2}$ molecules remaining after a fluence f. The column density is determined using Eq. \ref{Eq.3},
\begin{equation}
    N = \frac{\int \tau_{\nu} \rm d\nu}{A}
    \label{Eq.3}
,\end{equation}
where \emph{N} is the column density (molecules~cm$^{-2}$), \emph{A} is the band strength (cm~molecule$^{-1}$), and \emph{$\int$$\tau_{\nu}$\rm d$\nu$} is the integrated band in optical depth derived from the absorbance. The band strength \emph{A} can be calculated from the integrated extinction coefficient using Eq. \ref{Eq.4},
\begin{equation}
    A = \frac{\int \alpha \rm d\nu}{\rm {\rho_N}}
    \label{Eq.4}
,\end{equation}
where \emph{$\int$$\alpha$\rm d$\nu$} is the integrated extinction coefficient (cm$^{-1}$) and $\rm {\rho_N}$ is the number density (molecules cm$^{-3}$). Finally, by combining  Eq. \ref{Eq.3} and Eq. \ref{Eq.4}, we obtained  Eq. \ref{Eq.5}:
\begin{equation}
    N = \frac{\int \tau_{\nu} \rm d\nu \times \rho_N}{\int \alpha \rm d\nu}
    \label{Eq.5}
.\end{equation}
For our calculation we used an integrated extinction coefficient value of 3.9~$\times$~10$^{3}$ cm$^{-2}$ for the 2013 cm$^{-1}$ band calculated for the crystalline diacetylene at 70~K by \citet{Khannaetal.1988} and a $\rm {\rho_N}$ of 9.2~$\times$~10$^{21}$ molecules~cm$^{-3}$. The value of the number density is the more important source of uncertainties for this calculation and this value is derived from the mass density using Avogadro’s number, and the molar mass of diacetylene (50.06~g~mol$^{-1}$). However, the mass density of crystalline diacetylene is not available in the literature. Thus, we assumed a mass density of 0.76~g~cm$^{-3}$ measured for crystalline acetylene at 131~K \citep{Hudsonetal.2014, McMullanetal1992}. Figure \ref{Figure 5} presents the evolution of the consumed C$_{4}$H$_{2}$ in column density as a function of the photon-fluence for both experiments performed at JPL and PIIM. As explained in Sect. \ref{Infrared spectroscopy}, a doubled pathlength of absorption due to the reflection-absorption technique used at PIIM was factored in when calculating the column density for the PIIM experiments. We obtained that 2.6~$\times$~10$^{17}$~molecule~cm$^{-2}$ of C$_{4}$H$_{2}$ were consumed for a photon-fluence of 3.9~$\times$~10$^{21}$~photons~cm$^{-2}$ at 355~nm, while 3.4~$\times$~10$^{17}$~molecule~cm$^{-2}$ of C$_{4}$H$_{2}$ were consumed for a fluence of 4.5~$\times$~10$^{20}$~photons~cm$^{-2}$ at $\lambda$~>~300~nm. These results indicate that a polychromatic photon flux covering a broader wavelength range longward of 300~nm substantially increases the kinetics of the diacetylene photolysis, leading to a slightly higher number of diacetylene molecules consumed with a photon fluence that is ten times lower compared to irradiation performed at 355~nm. We can make the hypothesis that this difference is due to a significant increase in the absorbance of diacetylene between 350 and 300~nm (Fig. \ref{Figure 3}).

\begin{figure}[h]
    \centering
    \includegraphics [width=\hsize] {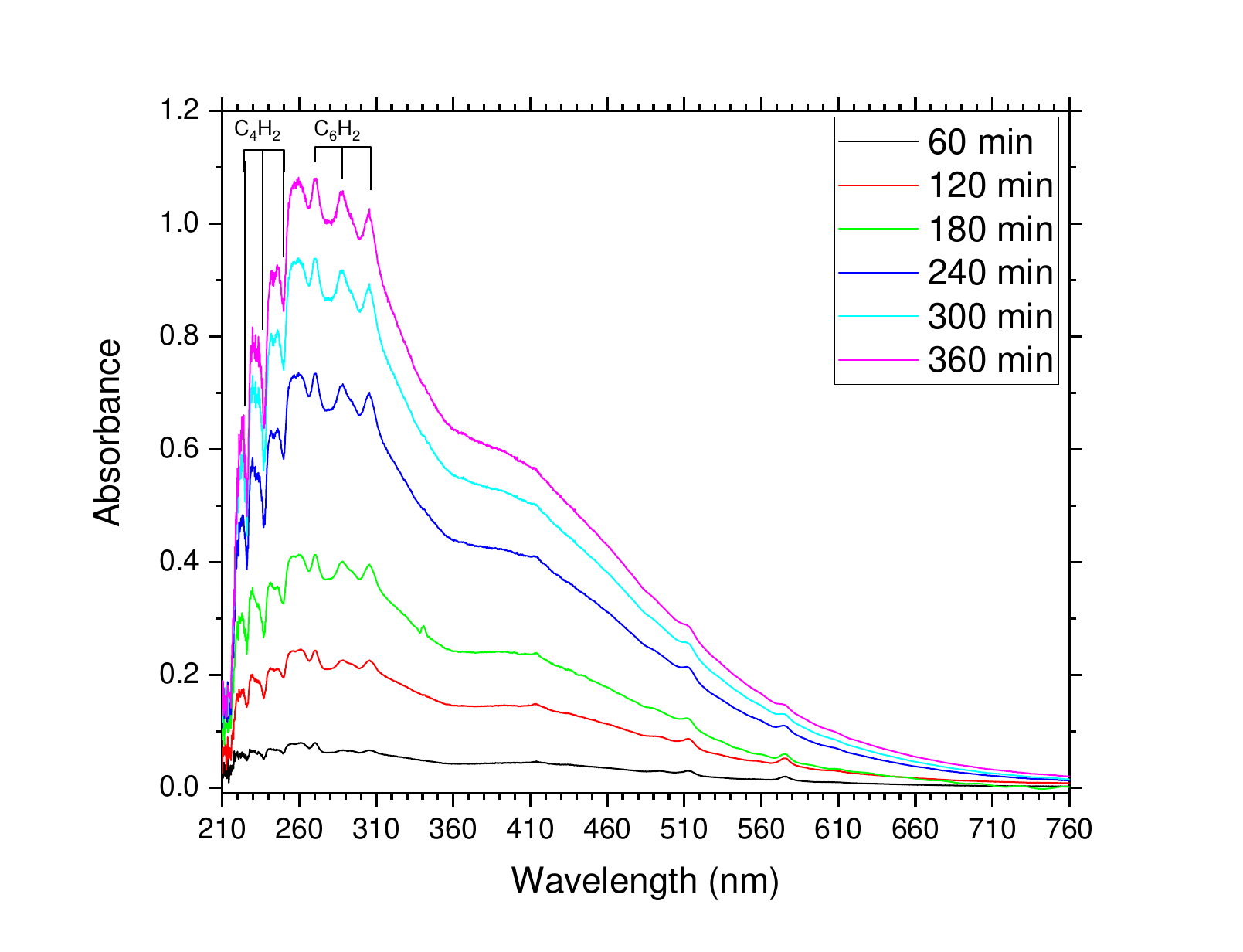}
    \caption {Differential absorbance (A$_{t}$ – A$_{t=0}$) in the UV-Vis domain for the different time of irradiation at 355~nm. Assignments of some of the absorption bands are shown.}
    \label{Figure 6}%
\end{figure}

The UV-Vis spectra obtained during the experiments at JPL are presented in Fig. \ref{Figure 6}; they provide complementary information on the photochemistry of C$_{4}$H$_{2}$ upon irradiation at 355~nm. We observed the diminution of the diacetylene absorption bands after 60~min of irradiation and then after each successive step of irradiation, in agreement with the IR measurements. In addition, new absorption bands were observed at longer wavelengths: 270, 287, 306, 414, 513, and 575~nm. In addition to these individual bands, we observed an increase in continuum from 260 to 700~nm after each irradiation. A similar absorption continuum was also observed during previous experiments on C$_{4}$N$_{2}$ ice \citep{Gudipatietal.2013} and can be attributed to the growth of a polymeric organic material in the ice. The new bands can be attributed to the production of new species. The bands at 270, 287, and~306 nm can be tentatively assigned to C$_{6}$H$_{2}$. In the gas phase C$_{6}$H$_{2}$ has strong absorption bands located at 260, 275, and 293~nm \citep{Jollyetal.2008, Shindoetal.2003}. In the solid phase we observed a shift of the absorption bands of C$_{6}$H$_{2}$ toward longer wavelengths by about 10~nm compared to the gas phase, as also observed for C$_{4}$H$_{2}$. We did not observe absorption bands of C$_{6}$H$_{2}$ in the infrared, but the C$_{6}$H$_{2}$ absorption bands are most likely buried under the  C$_{4}$H$_{2}$ bands (particularly in the C-H stretch region), which prevents its identification by IR spectroscopy. Other absorption bands at longer wavelengths could not be attributed to a specific compound. Nevertheless, polycyclic aromatic compounds (PAHs) are good candidates to explain these absorption bands because the formation of PAHs has been observed in another experimental study after the irradiation of acetylene ice (C$_{2}$H$_{2}$) with low-energy electrons \citep{Abplanalpetal.2019}. Further, several PAHs, such as phenanthrene (C$_{14}$H$_{10}$), perylene (C$_{20}$H$_{12}$) \citep{Salamaetal.2011}, or larger PAHs \citep{Ruiterkampetal2002}, have a strong absorption in the UV at wavelengths shorter than 300~nm and additional weaker absorption bands at longer wavelengths, similar to those observed in Fig. \ref{Figure 6}. However, the fundamental transition of these PAHs overlaps with the  absorption bands of other volatiles, such as C$_{6}$H$_{2}$, and  with the absorption continuum of the polymer. The UV-Vis spectra alone are not specific enough to identify, if formed during the experiments, a single PAH among many other photoproducts. Future work employing other analytical techniques such as chromatography mass spectrometry could provide a better identification of these molecules.

%-----------------------------------------------------------------
\subsection{Temperature-programmed desorption (TPD) of the ice after irradiation}\label{Temperature-programmed desorption (TPD) of the ice after irradiation}

\begin{figure}[h]
    \centering
    \includegraphics [width=\hsize] {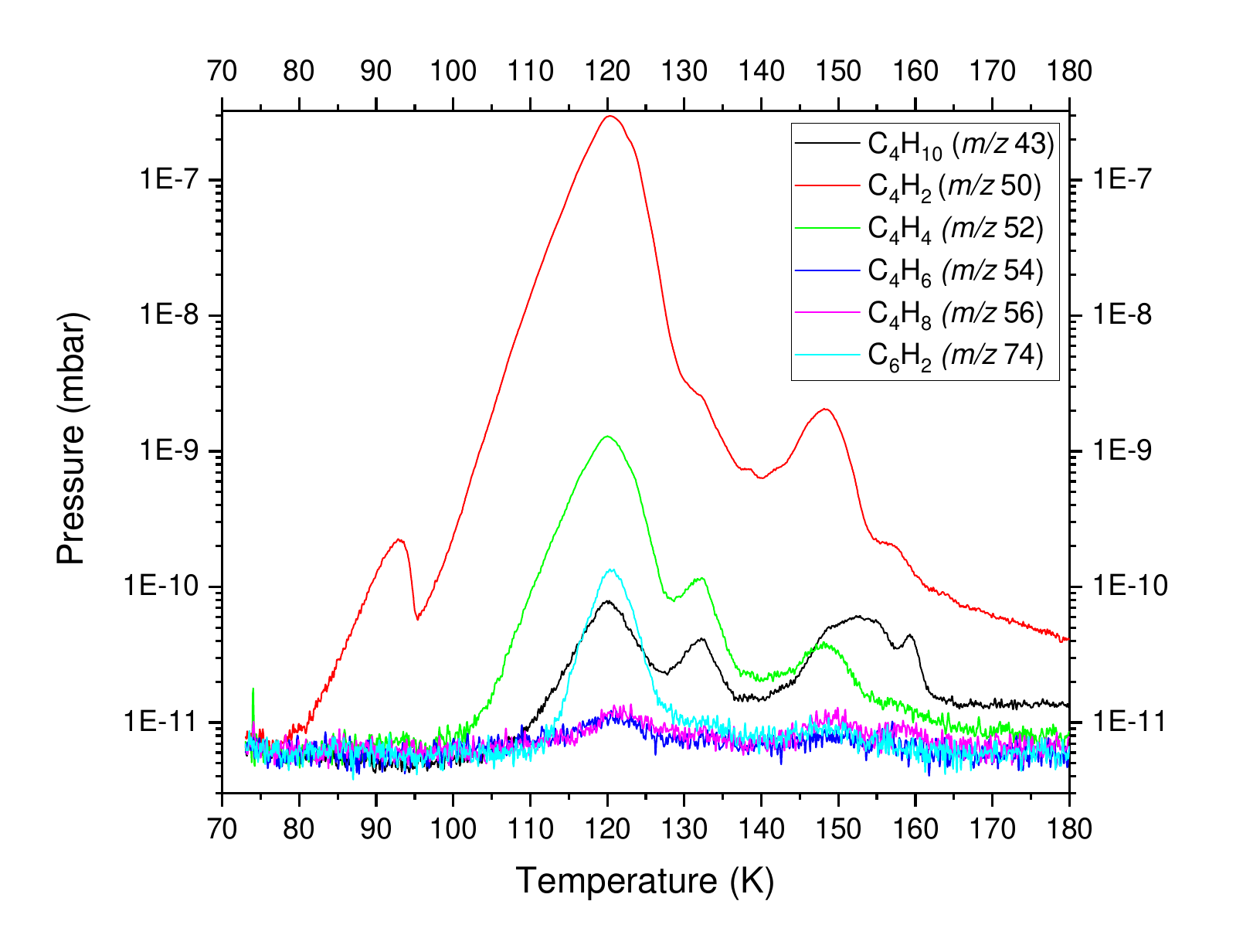}
    \caption {Temperature desorption profile of \textit{m/z} 43 (C$_{3}$H$_{7}^{+}$, main fragment of C$_{4}$H$_{10}$), 50 (C$_{4}$H$_{2}^{+}$), 52 (C$_{4}$H$_{4}^{+}$), 54 (C$_{4}$H$_{6}^{+}$), 56 (C$_{4}$H$_{8}^{+}$), and 74 (C$_{6}$H$_{2}^{+}$) after irradiation of C$_{4}$H$_{2}$ with a laser at 355~nm. The sample was heated (1~K~min$^{-1}$) from 70 to 180~K.}
    \label{Figure 7}%
\end{figure}

After the irradiation of the diacetylene ice at JPL, the sample was warmed under vacuum from 70 to 180~K with a ramp of 1~K~min$^{-1}$ to sublimate all the volatile species formed during irradiation (180~K is the lowest temperature to observe the sublimation of all volatiles). Desorption of several species was tracked as a function of the sample temperature using in situ mass spectrometry. After preliminary experiments in which we recorded mass spectra from \textit{m/z} 1 to 200 during TPD experiments, we decided to track the thermal desorption of C$_{4}$H$_{2}$ (C$_{4}$H$_{2}^{+}$, \textit{m/z} 50) and its hydrogenation products detected by IR spectroscopy, C$_{4}$H$_{4}$ (C$_{4}$H$_{4}^{+}$, \textit{m/z} 52). We also monitored the thermal desorption of other, more saturated C4 species: butyne C$_{4}$H$_{6}$ (C$_{4}$H$_{6}^{+}$, \textit{m/z} 54), butene C$_{4}$H$_{8}$(C$_{4}$H$_{8}^{+}$, \textit{m/z} 56), and butane (C$_{4}$H$_{10}$) using its major fragmentation peak (C$_{3}$H$_{7}^{+}$, \textit{m/z} 43) as the proxy. Finally, we tracked the thermal desorption of C$_{6}$H$_{2}$ (C$_{6}$H$_{2}^{+}$, \textit{m/z} 74), which was identified using UV-Vis spectroscopy. We did not observe an increase in the intensity at other masses, except for those corresponding to fragments of these molecules. Figure \ref{Figure 7} presents the temperature-dependent desorption profile for \textit{m/z} 43, 50, 52, 54, 56, and 74.

The C$_{4}$H$_{2}$ ice remaining after the irradiation is the most abundant species detected during the TPD. Its desorption started at $\sim$80~K with a first peak at 90~K, while the maximum of desorption occurred at $\sim$120~K;  a last peak is visible at $\sim$150~K. In a previous study crystalline C$_{4}$H$_{2}$ was found to start desorbing at $\sim$100~K with maxima at 120 and $\sim$150~K \citep{Zhouetal.2009}. Therefore, our results agree with this previous study. The ion signal at \textit{m/z} 52 starts to increase at 100~K with a maximum at 120~K, and can be attributed to the desorption of C$_{4}$H$_{4}$, which was also detected using IR spectroscopy. Finally, ion signals at \textit{m/z} 43, 54, 56, and 74 started to increase at 110~K with maxima at 120~K, coinciding with maximum desorption of C$_{4}$H$_{2}$. C$_{4}$H$_{4}$ is the main volatile product of the  C$_{4}$H$_{2}$ photochemistry detected during the TPD followed by C$_{6}$H$_{2}$ and C$_{4}$H$_{10}$ (based on C$_{3}$H$_{7}^{+}$, the  C$_{4}$H$_{10}$ main fragment). Ion signal variations at \textit{m/z} 54 and 56 were very low, showing that these two species were produced in a negligible amount compared to the three other hydrocarbons produced. It is likely that C$_{4}$H$_{6}$ and C$_{4}$H$_{8}$ are highly reactive compared to the other products. 

%-----------------------------------------------------------------
\subsection{Organic residues spectra}\label{Organic residues spectra}

\begin{figure}[h]
    \centering
    \includegraphics [width=\hsize] {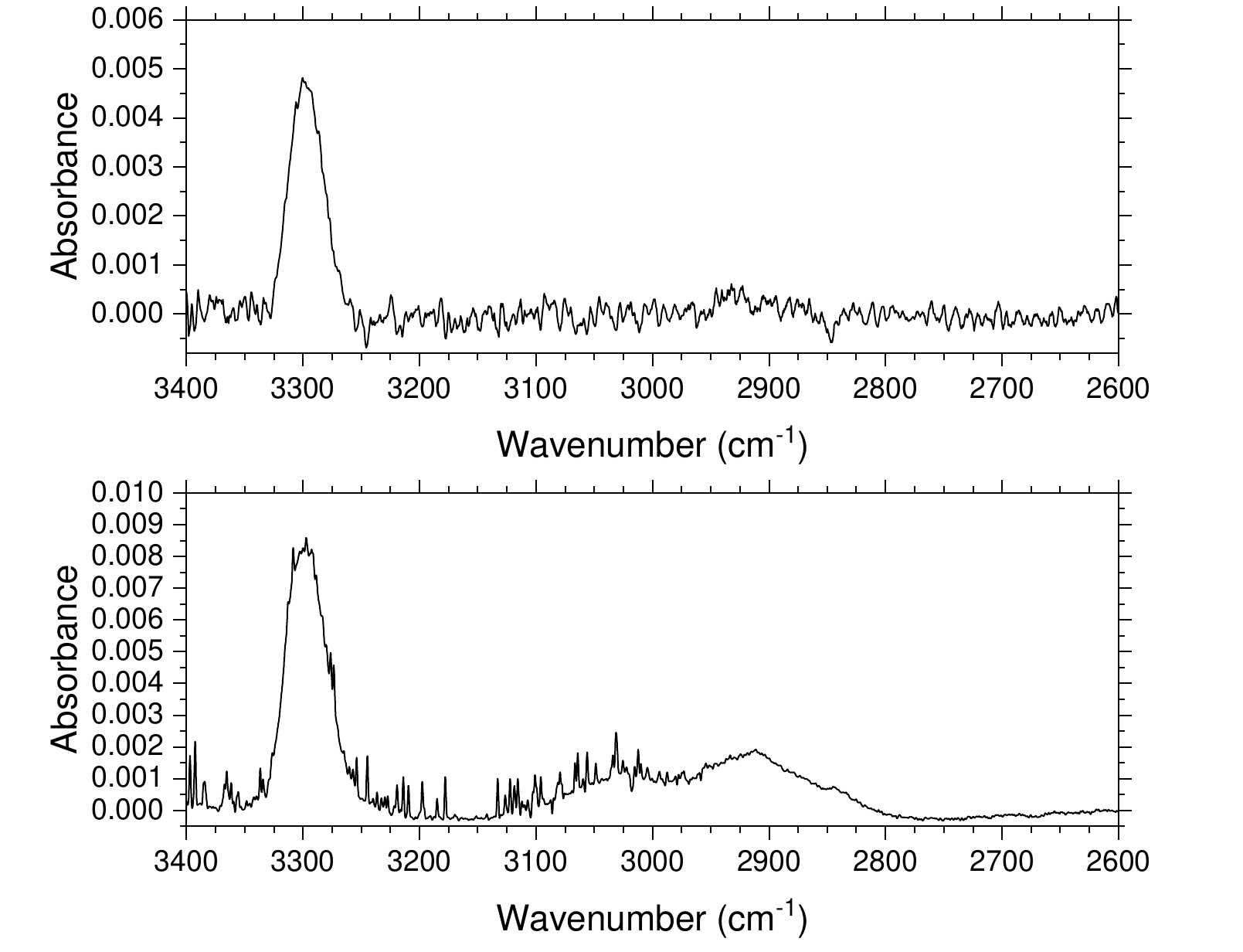}
    \caption {Infrared spectra obtained from non-volatile residue subsequent to C$_{4}$H$_{2}$ photolysis. Top: IR absorption spectrum of the organic polymer produced during the JPL experiment. Bottom: IR absorption spectrum of the organic polymer produced during the PIIM experiment. The spectra were recorded at ambient temperature at the end of the experiments.}
    \label{Figure 8}%
\end{figure}

At the end of the TPD run, each sample was further warmed to room temperature. Figure \ref{Figure 8} presents the IR spectra of the organic residues remaining, on the sapphire window (top) in the JPL studies and on the copper surface (bottom) in the PIIM studies, after the sublimation of the ices at room temperature under vacuum. These residues are lightly colored (brownish).

Both IR spectra (JPL and PIIM) show an intense absorption band $\sim$3300~cm$^{-1}$, as illustrated in Fig. \ref{Figure 8}, attributed to stretching $\equiv$CH. The position and shape of this absorption band of the polymer are similar for  the two different C$_{4}$H$_{2}$ ice photochemistry studies (355~nm laser irradiation at JPL and $\lambda$ > 300~nm Hg lamp irradiation at PIIM). However, the spectra of the organic films generated at JPL and at PIIM (Fig. \ref{Figure 8}) also show significant differences between 3100 and 2800~cm$^{-1}$. The lower-frequency absorptions between 3100 and 3000~cm$^{-1}$ (due to alkene and aromatic C-H stretch) seen in the PIIM studies are consistently absent in the  JPL studies, indicating that the broad range of wavelengths used in the PIIM irradiation (unlike the single laser wavelength used during the JPL experiments) resulted in more unsaturated and aromatic hydrocarbons. 

\begin{figure}[h]
    \centering
    \includegraphics [width=\hsize] {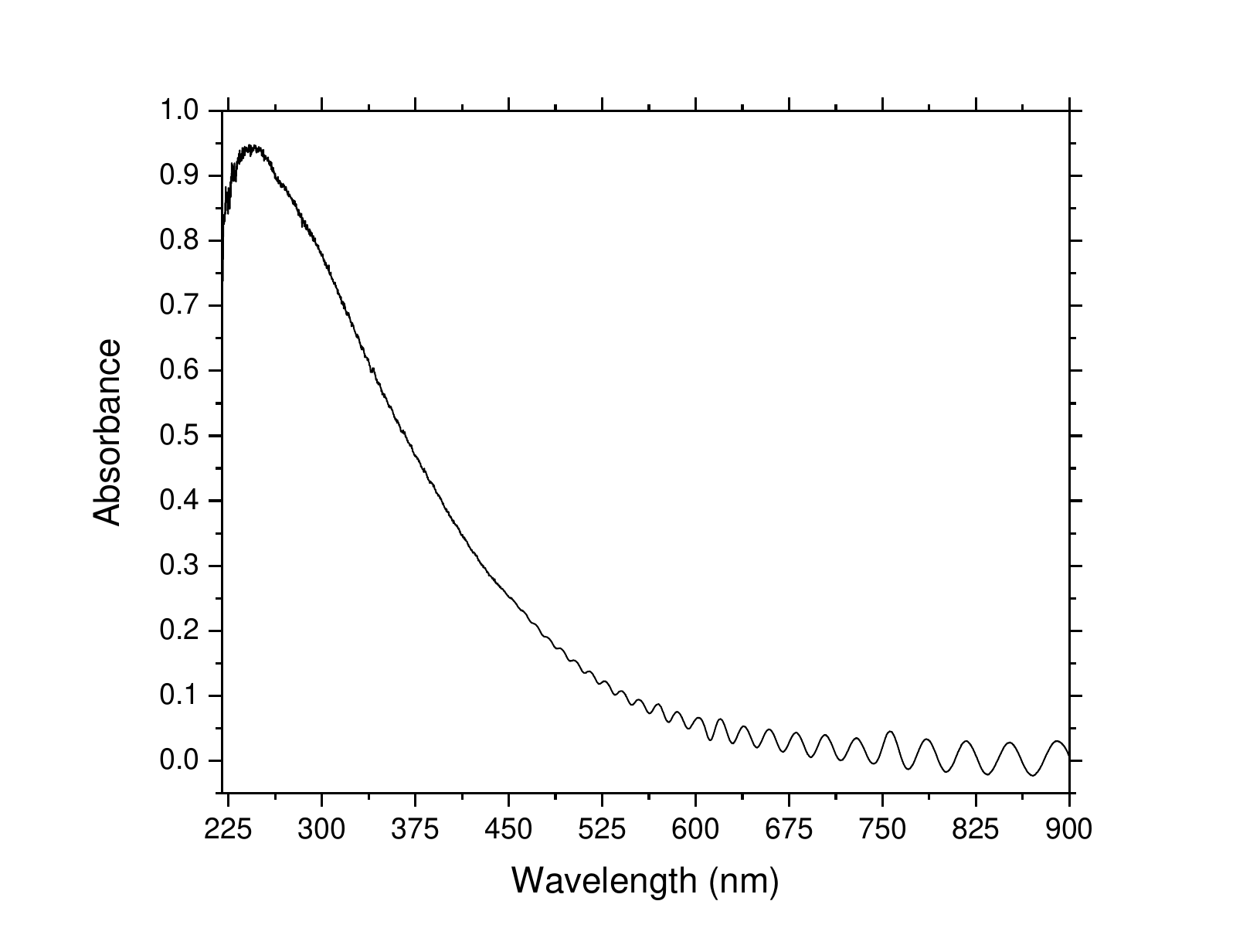}
    \caption {UV-Vis absorption spectrum of the organic polymer produced during the JPL experiment. The spectrum was recorded at ambient temperature at the end of the experiment.}
    \label{Figure 9}%
\end{figure}

In addition to the IR spectroscopy, the absorption spectrum of the polymer was recorded in the UV-Vis for the sample produced at JPL. The spectrum is presented in Fig. \ref{Figure 9}. At room temperature, the organic residue presents an intense absorption continuum in the UV, which decreases exponentially in the visible and extends up to 600~nm. Observations of interference fringes between 500 and 900~nm indicate that the polymer residue forms a nonscattering and homogeneous film on the sapphire window. The continuum absorption of the residue does not differ significantly from that observed at 70~K (Fig. \ref{Figure 6}), indicating that a large part of the polymer is formed during the irradiation of the ice. We observed at room temperature that all the absorption bands attributed previously to C$_{4}$H$_{2}$, C$_{6}$H$_{2}$, and tentatively to PAHs have disappeared, leaving only a large continuum. This indicates that most of these species are volatile at room temperature under vacuum, as illustrated by the detection of desorbed C$_{4}$ hydrocarbons and C$_{6}$H$_{2}$ by mass spectrometry during the TPD (Fig. \ref{Figure 7}). It is also likely that thermal reactions could occur between volatile compounds or between volatile compounds and the polymer during the warming up of the sample, modifying the composition of the polymer, which would be reflected in its UV-Vis absorption spectrum.

%-----------------------------------------------------------------
\section{Discussions}\label{Discussions}

%-----------------------------------------------------------------
\subsection{Chemical pathways for hydrocarbon formations}\label{Chemical pathways for hydrocarbons formations}

Our experiments point out that C$_{4}$H$_{2}$ is photochemically highly active in the solid phase, leading to the formation of C$_{6}$H$_{2}$ and hydrogenated products (C$_{4}$H$_{4}$, C$_{4}$H$_{6}$, C$_{4}$H$_{8}$, and C$_{4}$H$_{10}$). However, we did not find clear evidence of the formation of C$_{8}$H$_{2}$), which was observed previously in the gas phase irradiation of C$_{4}$H$_{2}$ \citep{Bandyetal.1992a,Bandyetal.1992b}. Although it is difficult to identify the reaction mechanism in our experimental conditions, based on literature data \citep{Bandyetal.1992a,Bandyetal.1992b}, we can try to propose chemical pathways for the formation of the different species observed. 

During the JPL experiments, irradiation of C$_{4}$H$_{2}$ with the laser at $\lambda$ = 355~nm (3.49~eV or $\sim$28169~cm$^{-1}$) can provide enough energy to excite C$_{4}$H$_{2}$ into the second lowest exited state $^{3}\Delta_{u}$, extended between 3.2 and 4.7~eV ($\sim$263~nm or $\sim$38023~cm$^{-1}$ and $\sim$400~nm or $\sim$25000~cm$^{-1}$) in the gas phase, as predicted by \citet{Allanetal.1984} and \citet{Bandyetal.1992b}. This also suggests that the JPL experiments could access the first triplet excited state $^{3}\Sigma_{u}^{+}$ centered at 3.2~eV ($\sim$387~nm or $\sim$25800~cm$^{-1}$), which should be involved in the photochemistry observed. Hence, the observed photochemistry must be mediated by the two lowest triplet states ($^{3}\Sigma_{u}^{+}$ or $^{3}\Delta_{u}$) reached through direct excitation from the ground singlet state, similar to the mechanism proposed for the C$_{4}$N$_{2}$ ice photochemistry \citep{Gudipatietal.2013}.

Likewise, for the experiments conducted at PIIM with $\lambda$ > 300~nm (E < 4.13~eV or $\sim$33333~cm$^{-1}$), we can access the UV absorption bands below 4.13~eV. As seen from Fig. \ref{Figure 3}, and as in the case of the JPL experiments, C$_{4}$H$_{2}$ can be also excited into the two first excited states $^{3}\Sigma_{u}^{+}$ or $^{3}\Delta_{u}$ \citep{Allanetal.1984} with the lamp and filter combination used at PIIM. However, these two kinds of irradiation ($\lambda$ = 355~nm and $\lambda$ > 300~nm) are not sufficient to excite the $^{1}\Sigma_{u}^{-}$ (4.8~eV or $\sim$38800~cm$^{-1}$) or $^{1}\Delta_{u}$ (5.1~eV or $\sim$40800~cm$^{-1}$), which are known to induce the C$_{4}$H + H and C$_{2}$H formation \citep{Bandyetal.1992b}. According to the work of \citet{Bandyetal.1992b} and more recently \citet{Vuitonetal2019}, at the wavelengths of our work at PIIM or JPL, photochemical reactions are supposed to be initiated from the metastable higher triplet excited state of C$_{4}$H$_{2}$, instead of the direct photolysis. These reaction channels are illustrated below. They can explain the formation of three compounds observed during our irradiation experiments namely C$_{4}$H$_{4}$, C$_{4}$H$_{6}$, and C$_{6}$H$_{2}$:

\begin{equation}
\begin{split}
\rm C_{4}H_{2}^{*}(^{3}\Delta_{u}) + C_{4}H_{2} &\longrightarrow \rm C_{8}H_{3} + H \\
\rm &\longrightarrow \rm C_{6}H_{2} + C_{2}H_{2} \\
\rm &\longrightarrow \rm C_{8}H_{2} + H_{2} 
    \label{Eq.6}
\end{split}
,\end{equation}

\begin{equation}
\begin{split}
\rm C_{4}H_{2}^{*}(^{3}\Delta_{u}) + 2H &\longrightarrow \rm C_{4}H_{4}
    \label{Eq.7}
\end{split}
,\end{equation}

\begin{equation}
\begin{split}
\rm C_{4}H_{4}^{*} + 2H &\longrightarrow \rm C_{4}H_{6}
    \label{Eq.8}
\end{split}
.\end{equation}
\citet{Bandyetal.1992b} proposed additional routes for the formation of secondary species via the reaction of C$_{4}$H$_{2}$ with primary products, as illustrated by reaction \ref{Eq.9}:
\begin{equation}
\begin{split}
\rm C_{8}H_{2} + C_{4}H_{2} &\longrightarrow \rm 2C_{6}H_{2}
    \label{Eq.9}
\end{split}
.\end{equation}
This secondary reaction could explain the nondetection of C$_{8}$H$_{2}$ at the end of our experiment.  \citet{Bandyetal.1992b} explained that the C$_{6}$H$_{2}$ formation induces the production of C$_{2}$H$_{2}$, as previously illustrated in Reaction \ref{Eq.6}. In addition, work from \citet{Glickeretal.1987} suggests that C$_{4}$H$_{2}$ photopolymerization occurred via molecular mechanisms, free radicals being detected only at $\lambda$ < 180~nm, so it seems highly probable that in both our experiments at JPL and PIIM (irradiations performed at $\lambda$ > 300~nm and $\lambda$ = 355~nm) the photochemistry occurred through the metastable triplet excited state pathway and not through the involvement of singlet excited states. Similarly, the formation of C$_{4}$H$_{4}$ seems to be driven by the reaction of C$_{4}$H$_{2}^{*}$ and H in solid phase \citep{Arringtonetal.1998}, as depicted by (Reaction \ref{Eq.7}). Based on the above analysis, we conclude that C$_{4}$H$_{2}$ ice photochemistry is driven by the direct photoexcitation into the triplet excited states, as previously observed with  C$_{4}$N$_{2}$ ice \citep{Gudipatietal.2013}. At least, C$_{4}$H$_{8}$ and C$_{4}$H$_{10}$ could be formed by successive addition of 2H on  C$_{4}$H$_{6}$ and  C$_{4}$H$_{8}$. Such photochemistry is accessible only in the condensed phase, where the spin-selection rules are relaxed for the absorption of a photon by a molecule with other molecules, making ground-state singlet to excited state triplet transitions weakly allowed. As a result, longer-wavelength photons can be absorbed through these transitions and can initiate photochemical reactions.  

%-----------------------------------------------------------------
\subsection{Implications for Titan’s atmospheric and surface chemistry}\label{Implications to Titan’s atmospheric and surface chemistry}

Our experimental results reveal that the exposure of diacetylene ice particles to near-UV photons ($\lambda$ > 300~nm) could drive a complex organic chemistry in Titan’s lower atmosphere. However, our results point out that the kinetics of the diacetylene photochemistry vary significantly as a function of the wavelength of the photons used for the irradiation. Our experiments at JPL show that a significant amount of photon fluence (3.9~$\times$~10$^{21}$~photons~cm$^{-2}$) is necessary with a monochromatic laser irradiation at 355~nm to obtain a moderate depletion of diacetylene ice. However, with a polychromatic photon flux (which is expected on Titan) covering wavelengths down to 300~nm, the kinetic of the diacetylene photolysis drastically increases leading to similar diacetylene molecules consumption with a fluence (4.5~$\times$~10$^{20}$~photons~cm$^{-2}$) ten times lower compared to the monochromatic experiments. We can qualitatively understand this difference as being due to a significant increase in the absorbance of diacetylene between 350 and 300~nm (Fig. \ref{Figure 3}) as well as broadband polychromatic lamp versus monochromatic laser as the photolysis source. Nevertheless, the flux of photons used in our experiments is orders of magnitude larger than that reaching the stratosphere and the troposphere of Titan. To enable a more rigorous quantification of our laboratory data in regard to Titan’s conditions, it is necessary to have better modeling studies that estimate photon fluxes at different wavelengths below 350~nm and at different altitudes in Titan’s atmosphere. This  estimation would allow us to quantify the fraction of the diacetylene ice that could be photolyzed in the atmosphere of Titan during the ice particles sedimentation before reaching the surface.

Moreover, this would allow us to quantify the impact of the solid-phase photochemistry for the composition of Titan’s lower atmosphere. This solid-phase chemistry could act as a sink for volatile species such as C$_{4}$H$_{2}$ in Titan's lower atmosphere and provide a source for other volatile hydrocarbons such as C$_{6}$H$_{2}$, although these products may be trapped in the ice phase depending on the pressure and temperature. We also expect other atmospheric molecules to precipitate and accrete onto the ice at lower altitudes all the way to the surface. Other major atmospheric molecules of Titan such as HCN and other nitriles, which are known constituents of Titan’s ice clouds \citep{Andersonetal.2011, Andersonetal.2010, deKoketal.2014, Samuelsonetal.1997}, could undergo similar reactivity in the solid phase. An analysis of the spectral signatures of ice clouds suggests that those ice particles could be made of co-condensed molecular species \citep{Andersonetal.2018a}. In the case of the co-condensation of a hydrocarbon such as C$_{4}$H$_{2}$ and a nitrogen-bearing species such as HCN or HC$_{3}$N, solid-phase reactivity likely leads to a coupling of the nitrogen and hydrocarbon chemistry and may initiate the formation of nitrogen containing complex organics, including N-PAHs and possibly biologically important molecules such as adenine (C$_{5}$H$_{5}$N$_{5}$). Overall, this solid-phase reactivity plays an important role in determining the composition of the organic material formed in the atmosphere and that will ultimately contribute to the complex organic molecular, ice, and aerosol diversity on Titan’s surface, whose composition will be studied by the future Dragonfly mission.

%-----------------------------------------------------------------
\section{Conclusions}\label{Conclusions}
Titan's atmosphere is enriched with simple hydrocarbons and nitriles. Some of the most abundant molecules are detected both in the gas phase and in the condensed phase in the atmosphere. Diacetylene is one exception, which is detected in the gas phase but not in the condensed phase. Non-detection of C$_{4}$H$_{2}$ in the atmospheric ice clouds as well as on the surface of Titan has been a puzzle, for which the results of our laboratory experiments provide a potential rationale.
   \begin{enumerate}
      \item We studied experimentally the photochemistry of solid diacetylene at 70~K under irradiation with photons at wavelengths longer than 300~nm, closely simulating the radiations received by ice particles in Titan’s atmosphere. Our experimental results have shown that diacetylene ice is very reactive under these conditions. A total of  2.6~$\times$~10$^{17}$~molecules~cm$^{-2}$ of C$_{4}$H$_{2}$ were consumed after a photon-fluence of 3.9~$\times$~10$^{21}$~photons~cm$^{-2}$ at 355~nm while 3.4~$\times$~10$^{17}$~molecules~cm$^{-2}$ of C$_{4}$H$_{2}$ were consumed after a photon-fluence of 4.5~$\times$~10$^{20}$~photons~cm$^{-2}$ at $\lambda$ > 300~nm, indicating increasing photochemistry yields with decreasing photon wavelengths between 355~nm and 300~nm.
      \item We analyzed the evolution of the composition of the solid phase after irradiation using IR and UV-Vis spectroscopy. In addition, we used TPD to analyze the composition of the desorbed species using in situ mass spectrometry. Our results show  that the diacetylene photochemistry follows different pathways leading to the formation of different volatile species as well as an organic residue. First, we observed hydrogenation of C$_{4}$H$_{2}$ to form other C$_{4}$H$_{2}$n (n = 2-5) hydrocarbons. The main photoproduct was C$_{4}$H$_{4}$ and then C$_{4}$H$_{10}$, while C$_{4}$H$_{6}$ and C$_{4}$H$_{8}$ were produced in a negligible amount. Second, the photochemistry of C$_{4}$H$_{2}$ ice also resulted in the production of larger and highly polymerizable hydrocarbons such as C$_{6}$H$_{2}$, and presumably aromatics compounds such as PAHs. 
      \item Finally, we observed the very efficient formation of a tholin-like organic polymer that is stable at room temperature, which has a strong absorption continuum at UV-Vis wavelengths, monotonically decreasing from 200 to 600~nm. Beyond 600~nm no absorption was detected, indicating that the photopolymer could be photochemically activated with photons at wavelengths between 250 and 600~nm either in Titan’s troposphere and stratosphere or on the surface. These results agree with our earlier work on other tholins, indicating that this is the general behavior of Titan’s organic aerosol analogs.
      \item Although diacetylene ice has not been observed yet on Titan, our experimental results highlight the importance that this compound could have for the ice photochemistry in Titan’s troposphere and stratosphere. A significant amount of C$_{4}$H$_{2}$ ice could be transformed into other volatile molecules, and into  organic polymers, and so provide a significant loss process for diacetylene. It is likely that in the case of co-condensation of two or more volatiles, C$_{4}$H$_{2}$ could trigger the photochemistry through reaction with other hydrocarbons or nitriles, initiating a more complex chemistry in the solid phase. 
   \end{enumerate}

\begin{acknowledgements}
      Some of this work was carried out at the Jet Propulsion Laboratory, California Institute of Technology under a contract with the National Aeronautics and Space Administration. The JPL part of the work was supported by the NASA Solar System Workings and New Frontiers Data Analysis programs. The work carried out at the Physique des Interactions Ioniques et Moléculaires (PIIM) laboratory, Aix Marseille Université, was supported by the Programme National de Planétologie (PNP).
\end{acknowledgements}

\bibliography{48658corr} % your references Yourfile.bib
\bibliographystyle{aa} % style aa.bst

\end{document}